\documentstyle[preprint,aps,epsfig]{revtex}
\def\mycomm#1{\hfill\break\strut\kern-3em{\tt ====> #1}\hfill\break}
\def\mycommNL#1{\strut\kern-3em{\tt ====> #1}\hfill\break}

\newcommand{\eqref}[1]{(\ref{#1})}   

\makeatletter
\def\hlinewd#1{\noalign{\ifnum0=`}\fi
\hrule \@height #1 \futurelet \reserved@a\@xhline}
\def\hwhiteline{\noalign
{\ifnum0=`}\fi\hrule
\@height 0pt\vskip 1.0ex\futurelet \reserved@a\@xhline}
\makeatother
\def\gray{\special{ps: 0.40 setgray}}
\def\black{\special{ps: 0.0 setgray}}

\newcommand{\mydraft}{
\newcount\timecount
\newcount\hours \newcount\minutes  \newcount\temp \newcount\pmhours

\hours = \time
\divide\hours by 60
\temp = \hours
\multiply\temp by 60
\minutes = \time
\advance\minutes by -\temp
\def\hour{\the\hours}
\def\minute{\ifnum\minutes<10 0\the\minutes
    \else\the\minutes\fi}
\def\clock{
\ifnum\hours=0 12:\minute\ AM
\else\ifnum\hours<12 \hour:\minute\ AM
\else\ifnum\hours=12 12:\minute\ PM
    \else\ifnum\hours>12
     \pmhours=\hours
     \advance\pmhours by -12
     \the\pmhours:\minute\ PM
     \fi
    \fi
\fi
\fi
}
\def\fullclock{\hour:\minute}
\begin{centering}
\gray
\font\Hugett  =cmtt12 scaled\magstep3
\hbox{\kern-3em\Hugett Draft:\today,\clock}
\black
\end{centering}
\vskip -1.7cm
$\phantom{a}$
} 

\def\beq#1{\begin{equation} \label{#1}}
\def\eeq{\end{equation}}

\newskip\humongous \humongous=0pt plus 1000pt minus 1000pt

\newif\ifdtup

\begin{document}
{\tighten
 \preprint {\vbox{
  \hbox{$\phantom{aaa}$}
  \vskip-1.5cm
\hbox{Cavendish-HEP-04/08}
\hbox{TAUP 2763-04}
\hbox{WIS/04/04-JAN-DPP}
\hbox{ANL-HEP-PR-04-5} 
}}

\title{A Mass Inequality for the $\Xi^*$ and $\Theta^+$ Pentaquarks}

\author{Marek Karliner\,$^{a,b}$\thanks{e-mail: \tt marek@proton.tau.ac.il}
\\
and
\\
Harry J. Lipkin\,$^{b,c}$\thanks{e-mail: \tt
ftlipkin@clever.weizmann.ac.il} }
\address{ \vbox{\vskip 0.truecm}
$^a\;$Cavendish Laboratory\\
Cambridge University, England;\\
and\\
$^b\;$School of Physics and Astronomy \\
Raymond and Beverly Sackler Faculty of Exact Sciences \\
Tel Aviv University, Tel Aviv, Israel\\
\vbox{\vskip 0.0truecm}
$^c\;$Department of Particle Physics \\
Weizmann Institute of Science, Rehovot 76100, Israel \\
and\\
High Energy Physics Division, Argonne National Laboratory \\
Argonne, IL 60439-4815, USA\\
}

\maketitle
\begin{abstract}%
We derive an upper bound on the mass difference between the 
$\Xi^*$ and $\Theta^+$ pentaquarks which are the manifestly exotic 
members of the $SU(3)_f$ antidecuplet. The derivation is
based on simple assumptions about $SU(3)_f$ symmetry breaking and uses
the standard quantum mechanical variational method. The resulting 
rather robust bound is more than 20 MeV below the experimentally reported 
$\Xi^*-\Theta^+$ mass difference, emphasizing the need for confirmation of
the experimental mass values and placing strong constraints on quark 
models of the pentaquark structure.
 \end{abstract}%
\vfill \eject

\section{General arguments giving bounds on pentaquark masses}

There is now a general search for pentaquarks which are related to the 
$\Theta^+$ \cite{Theta,others}
by changes in flavor of one or more constituents. 

The recently observed large mass difference of $\approx 320$ MeV between  the
observed $\Xi^*$ \cite{Xi,Fischer:2004qb}
state and the  $\Theta^+$ places serious constraints on any
model. If these two states are members of the same $SU(3)$ 
antidecuplet \cite{antidec,DPP,Weigel,Prasz,Ellis:2004uz}, the mass
difference must be due to $SU(3)$ symmetry breaking. We show here using a nearly
model-independent variational principle that it is very difficult to fit this
mass difference with any simple model \cite{NewPenta,JW,JM,JW2}
for symmetry breaking.

We begin with the assumption that the exact wave functions for both the 
$\Theta^+$ and $\Xi^{*--}$ are eigenfunctions of the same QCD Hamiltonian and
that in the $SU(3)$ symmetry limit the masses of the two states are equal  and
that their wave functions differ only by the interchange of $u$ and $s$
flavors. We can therefore create a variational trial wave function for the 
$\Xi^{*--}$ by operating on the exact
wave function for the $\Theta^+$ with the $SU(3)$ transformation that
interchanges the $u$ and $s$ flavors. Although we do not know these exact
wave functions and do not know how the QCD Hamiltonian acts on them, we can
obtain interesting results for the mass difference by only making simple
assumptions about $SU(3)$ symmetry breaking.

The mass obtained with this wave function is an upper bound on 
the $\Xi^{*--}$ mass by the variational principle.
$\Theta^+$ contains one $\bar s$ antiquark and 4 light quarks, while
$\Xi^{*--}$ contains 2 strange quarks and 3 light (anti)quarks. 
Since the color field is
not changed by this flavor change with the same wave function, the only 
change in nearly all proposed models is one quark-mass change and a change in 
the color magnetic 
energy. 
\beq {vards}
M(\Xi^{*--}) \leq M(\Theta^+) + m_s - m_u + 
\langle \delta V_{hyp}(\bar s \rightarrow \bar u) \rangle_{\Theta^+} +
\langle \delta V_{hyp}(u \rightarrow s) \rangle_{\Theta^+} 
\end{equation}
where $\langle \delta V_{hyp}(\bar s \rightarrow \bar u) \rangle_{\Theta^+}$
and $\langle \delta V_{hyp}(u \rightarrow s) \rangle_{\Theta^+}$  denote the
changes in the color magnetic energy of the exact $\Theta^+$ wave function by
the replacements $\bar s \rightarrow \bar u$ and $u \rightarrow s$. 

Although we use the term ``color
magnetic energy", motivated by the color hyperfine interaction,
the description in terms of the  quantities 
$\delta V_{hyp}$ is considerably more general
and applies to any model in which
$SU(3)$ breaking is described by an additive quark mass term and a
flavor-dependent-spin dependent two-body quark-quark interaction denoted
by $V_{hyp}$\,, with its matrix elements determined by fitting ground state
baryon masses.

We now estimate these $SU(3)$ breaking effects by using inequalities which are 
are satisfied by most models.

The effective quark mass difference $m_s - m_u$
is given  by the $\Lambda$-nucleon mass difference 
\cite{NewPenta,SakhZel,Postcard,DGG}. This gives an upper bound satisfied by
nearly all models, 
\beq {sudif} m_s - m_u \leq M(\Lambda) - M(N) \end{equation}
Note that alternative values of $m_s - m_u$ commonly used are the baryon
decuplet mass splittings which give a value much less than the upper bound
(\ref {sudif}). Using such alternative values of $m_s - m_u$ would therefore
make the inequality \eqref{vards} even more stringent.

Since the color hyperfine interaction is proportional to the effective 
color magnetic moment, which in turn is inversely proportional to the quark
effective mass, the color-magnetic energy of the antiquark can only be 
lowered by the replacement $\bar s \rightarrow \bar u$. Therefore
\beq {hypqbar}
\langle \delta V_{hyp}(\bar s \rightarrow \bar u) \rangle_{\Theta^+} \leq 0 
\end{equation}

We now estimate  the maximum change in the
color-magnetic energy by changing the two $u$-quarks into $s$-quarks.

The color-magnetic energy between the two $u$-quarks is repulsive and is lowered
by the $u \rightarrow s$
transition. The color-magnetic energy of a $ud$ pair can be raised by the 
$u \rightarrow s$
transition.
We therefore estimate that
the maximum change in the color-magnetic energy is obtained by keeping the
two $u$-quarks apart, so that there is no change in their hyperfine energy,
and coupling each $u$-quark with one $d$-quark to get the maximum change in
the hyperfine energy for each.
 The maximum change to be expected is if it is bound to the $d$ quark
in a color antitriplet-spin-zero state, like the $ud$ pair in the $\Lambda$,
where there is no color-magnetic energy in the interaction with the strange 
quark. The change in the color-magnetic energy of a $ud$ pair in a color 
antitriplet-spin-zero state
produced by the $u \rightarrow s$ 
transition is obtainable from the $\Sigma - \Lambda$ mass difference. 

We choose a derivation here to minimize model dependence. The $ud$ pairs in the
$\Sigma$ and $\Lambda$  are both in color-antitriplet s-states and are thus
antisymmetric in space and color. They must therefore be symmetric in spin and
flavor and have isospins $I=1;S=1$ and $I=0;S=0$
respectively. In the flavor-symmetry limit, the $\Sigma$ and $\Lambda$
masses must be equal, since they are members of the same octet and the 
contribution of the color magnetic energies to the mass
difference of all pairs must cancel. Thus
\beq {lamsig}
M(\Sigma^-)-M(\Lambda) =
[\langle  V_{hyp}(ud) \rangle_{(S=1)} -
\langle  V_{hyp}(sd) \rangle_{(S=1)}] -
[\langle  V_{hyp}(ud) \rangle_{(S=0)} -
\langle  V_{hyp}(sd) \rangle_{(S=0)}] 
\end{equation}
This can be rewritten
\beq {lamsig2}
M(\Sigma^-)-M(\Lambda) =
(\kappa - 1)\cdot [
\langle  V_{hyp}(ud) \rangle_{(S=0)} -
\langle  V_{hyp}(sd) \rangle_{(S=0)}] 
\end{equation}
where 
\beq {lamsigkap}
\kappa =
{{\langle V_{hyp}(ud) \rangle_{(S=1)}}\over{\langle V_{hyp}(ud)\rangle_{(S=0)}}}
=
{{\langle V_{hyp}(sd) \rangle_{(S=1)}}\over{\langle V_{hyp}(sd)\rangle_{(S=0)}}}
\end{equation}
The maximum color-magnetic change occurs if both $u$ quarks have the
minimum color magnetic energy,
\beq {hypqq}
\langle \delta V_{hyp}(u \rightarrow s) \rangle_{\Theta^+} 
\leq 2 \cdot \langle \delta V_{hyp}(u \rightarrow s) \rangle_{ud(S=0)} 
={2\over{1-\kappa} }\cdot [M(\Sigma^-)-M(\Lambda)]
\end{equation}

For a hyperfine interaction proportional to $\vec \sigma_1 \cdot \vec 
\sigma_2$, $\kappa = -(1/3)$ and
we  get the inequality 
\beq {xistineq}
M(\Xi^{*--}) - M(\Theta^+) \leq M(\Lambda) - M(N) + 
{3\over 2}\cdot [M(\Sigma^-)-M(\Lambda)] = 299 \,{\rm MeV}   
\end{equation}
This is to be compared with with the mass difference of $\approx 330$
MeV between the reported $\Xi^*$ \cite{Xi} state 
 and the  $\Theta^+$  \cite{Theta,others}.
This slight but significant disagreement emphasizes the need
for confirming the experimental mass and places serious constraints on models.
This constraint will be seriously violated in any model where the $m_s - m_u$
quark mass difference is less than $M(\Lambda) - M(N)$ or the change in the
hyperfine energy including the hyperfine energy of the antiquark neglected in
eq. (\ref{xistineq}) is less than  $(3/2)\cdot[M(\Sigma)-M(\Lambda)]$. 

An alternative inequality is obtained if the parameter $\kappa$ is determined 
from experimental data, rather than the assumption that the  
interaction is proportional to $\vec \sigma_1 \cdot \vec 
\sigma_2$.
The $\Sigma^{*-}$ and $\Delta^{o}$ decuplet baryons differ by a 
$(u \rightarrow s)$ transition on the $\Delta^{o}$. This produces a mass change
of one quark mass difference and a change in the color magnetic energy of the
two $S=1$ $ud$ pairs in the $\Delta^o$. If we assume that the quark mass difference
and the color magnetic energy changes are the same in the baryon octet and
decuplet, we obtain 
 
\beq{sigdel}
M(\Sigma^{*-})-M(\Delta^o) = m_s - m_u + 
 2\langle \delta V_{hyp}(u \rightarrow s) \rangle_{ud(S=1)} 
\end{equation} 

\beq{sigdel2}
M(\Sigma^{*-})-M(\Delta^o)  = m_s - m_u + 
 2\langle \delta V_{hyp}(u \rightarrow s) \rangle_{ud(S=0)} 
 -2[M(\Sigma^-)-M(\Lambda)]
\end{equation}
where we have used eq. (\ref {lamsig}). 
Combining eqs. (\ref {vards}), (\ref {hypqbar}) and (\ref{sigdel2}) 
gives the inequality
\beq {xistineq2}
M(\Xi^{*--}) - M(\Theta^+) \leq M(\Sigma^{*-}) - M(\Delta^o) + 
2[M(\Sigma^-)-M(\Lambda)] = 316 \,{\rm MeV}   
\end{equation}
This uses the assumption that the color magnetic interactions are equal in the
octet and decuplet but does not use a theoretical relation between the singlet
and triplet splittings to determine the value of $\kappa$.

In order to consider analogous bounds for the other members of 
the $\Xi^*$ multiplet, one needs to take isospin breaking into account.
For baryons this is typically of order a few MeV, which is non-negligible
compared to the difference between \eqref{xistineq2}
and the $\Xi^*-\Theta^+$ splitting reported by experiments.

Our results
apply to any model with a flavor-dependent-spin-dependent
two-body
quark-quark interaction whose matrix elements satisfy 
eqs.~(\ref{hypqbar}-\ref{sigdel2})
with the
parameter $\kappa$ fixed by fitting the experimental data in eqs.
(\ref{sigdel}) and (\ref{xistineq2}).
Any small deviations from these conditions are easily tested by
applying our variational principle in any
model with a well defined $\Theta^+$ wave function and a well defined
prescription for flavor symmetry breaking.

Note that the bounds \eqref{sigdel2} and \eqref{sigdel} were derived 
with the assumption that the inequality (\ref{hypqbar}) 
is saturated. This is an extreme assumption which
implies no contribution from the
color-magnetic interaction of the $\bar s$ to the binding of the $\Theta^+$
and it may be unrealistic. It implies that the light quarks are
coupled to spin zero and do not interact with the antiquark. However, in any
simple model the  color-magnetic interaction of the $\bar s$ with a
quark system coupled to spin zero will polarize the quark system to spin one
and reduce the total energy. In this case the constraint (\ref{hypqbar}) will
be lowered and the disagreement of with experiment of the variational estimate
of $M(\Xi^{*--}) - M(\Theta^+)$ will become more severe. Thus the contribution
from the color-magnetic interaction of the $\bar s$ should be checked in any
model for the $\Theta^+$ and used to strengthen the variational estimate for
$M(\Xi^{*--}) - M(\Theta^+)$.

Small mass-dependent effects like the kinetic energy and possible spin-orbit
effects have been neglected. Kinetic energy differences are  probably included
in the effective mass difference, since $M(\Lambda) - M(N)$ includes the
difference in the kinetic energies of the $s$ quark in the $\Lambda$ and the
$u$ quark in the proton. If spin-orbit effects pull the $\Theta^+$ down more
than the $\Xi^{*--}$, then the mass gap  $M(\Xi^{*--}) - M(\Theta^+)$ can be
bigger than the above bounds (\ref {xistineq}) and (\ref {xistineq2}).  This
point has been addressed\cite{clodudek} with the interesting result  that the
$L\cdot S$ contributions for $\Theta^+$ and $\Xi^{*--}$ are expected to be very
similar and should only have a small affect on the bounds.

\section*{Note added -- April 2004}
Recently two experiments announced null results in search for $\Xi^{*--}$
\cite{WA89,CDF}. Since the production mechanism is yet unknown,
at this time it is not clear whether or not these results are
compatible with those of Ref.~\cite{Xi}.

\section*{Acknowledgements}

Discussions with J. J. Dudek and F. E. Close are gratefully acknowledged. 
The research of one of us (M.K.) was supported in part by a grant from the
United States-Israel Binational Science Foundation (BSF), Jerusalem.
The research of one of us (H.J.L.) was supported in part by the U.S. Department
of Energy, Division of High Energy Physics, Contract W-31-109-ENG-38.

%
\catcode`\@=11 
\def\references{
\ifpreprintsty \vskip 10ex
%
\hbox to\hsize{\hss \large \refname \hss }\else
\vskip 24pt \hrule width\hsize \relax \vskip 1.6cm \fi \list
{\@biblabel {\arabic {enumiv}}}
{\labelwidth \WidestRefLabelThusFar \labelsep 4pt \leftmargin \labelwidth
\advance \leftmargin \labelsep \ifdim \baselinestretch pt>1 pt
\parsep 4pt\relax \else \parsep 0pt\relax \fi \itemsep \parsep \usecounter
{enumiv}\let \p@enumiv \@empty \def \theenumiv {\arabic {enumiv}}}
\let \newblock \relax \sloppy
 \clubpenalty 4000\widowpenalty 4000 \sfcode `\.=1000\relax \ifpreprintsty
\else \small \fi}
\catcode`\@=12 

} 

\begin{thebibliography}{99}

\bibitem{Theta}
T.~Nakano {\it et al.}  [LEPS Collaboration],
Phys.\ Rev.\ Lett.\  {\bf 91} (2003) 012002
[arXiv:hep-ex/0301020].

\bibitem{others}
V.~V.~Barmin {\it et al.}  [DIANA Collaboration],
Phys.\ Atom.\ Nucl.\  {\bf 66} (2003) 1715
[Yad.\ Fiz.\  {\bf 66} (2003) 1763],
hep-ex/0304040;
S.~Stepanyan {\it et al.}  [CLAS Collaboration],
hep-ex/0307018.
%
J.~Barth {\it et al.}  [SAPHIR Collaboration],
hep-ex/0307083;
%
V.~Kubarovsky and S.~Stepanyan  and CLAS Collaboration,
hep-ex/0307088;
%
A.~E.~Asratyan, A.~G.~Dolgolenko and M.~A.~Kubantsev,
hep-ex/0309042.
%
V.~Kubarovsky {\it et al.}, [CLAS Collaboration],
hep-ex/0311046;
A. Airapetian {\it et al.}, [HERMES Collaboration],
arXiv:hep-ex/0312044;
S.~Chekanov, [ZEUS Collaboration],
{\tt http://www.desy.de/f/seminar/Chekanov.pdf};
%
R.~Togoo {\it et al.}, Proc. Mongolian Acad. Sci., {\bf 4} (2003) 2;
A.~Aleev {\it et al.}, [SVD Collaboration],
arXiv:hep-ex/0401024.

\bibitem{Xi}
C.~Alt {\it et al.}  [NA49 Collaboration],
arXiv:hep-ex/0310014.

\bibitem{Fischer:2004qb}
H.~G.~Fischer and S.~Wenig,
arXiv:hep-ex/0401014.

\bibitem{antidec}
P.~O.~Mazur, M.~A.~Nowak and M.~Prasza{\l}owicz,
Phys.\ Lett.\ B {\bf 147} (1984) 137;
A.~V.~Manohar,
Nucl.\ Phys.\ B {\bf 248} (1984) 19;
M.~Chemtob,
Nucl.\ Phys.\ B {\bf 256} (1985) 600;
S.~Jain and S.~R.~Wadia,
Nucl.\ Phys.\ B {\bf 258} (1985) 713;
M.~P.~Mattis and M.~Karliner,
Phys.\ Rev.\ D {\bf 31} (1985) 2833;
M.~Karliner and M.~P.~Mattis,
Phys.\ Rev.\ D {\bf 34} (1986) 1991;
M.~Prasza{\l}owicz, {\it Proc. of the Workshop on Skyrmions and Anomalies},
Krak\'ow, 1987, eds. M~Je\.zabek and M.~Prasza{\l}owicz (World
Scientific, Singapore, 1987), p.531;

\bibitem{DPP}
D.~Diakonov, V.~Petrov and M.~V.~Polyakov,
Z.\ Phys.\ A {\bf 359} (1997) 305
[arXiv:hep-ph/9703373].

\bibitem{Weigel}
H.~Weigel,
Eur.\ Phys.\ J.\ A {\bf 2} (1998) 391
[arXiv:hep-ph/9804260].

\bibitem{Prasz}
M.~Prasza{\l}owicz,
Phys.\ Lett.\ B {\bf 575} (2003) 234
[arXiv:hep-ph/0308114].

\bibitem{Ellis:2004uz}
J.~Ellis, M.~Karliner and M.~Prasza{\l}owicz,
arXiv:hep-ph/0401127.

\bibitem{NewPenta}
M. Karliner and H.J. Lipkin,
Phys.\ Lett.\ B {\bf 595} (2003) 249,
hep-ph/0307243, hep-ph/0307343.

\bibitem{JW}
R.~L.~Jaffe and F.~Wilczek,
Phys. Rev. Lett. {\bf 91} (2003) 232003,
[arXiv:hep-ph/0307341].

\bibitem{JM}
B.~K.~Jennings and K.~Maltman,
arXiv:hep-ph/0308286;
%
\bibitem{JW2}
R.~Jaffe and F.~Wilczek,
arXiv:hep-ph/0312369.

\bibitem{SakhZel}{
Ya.B. Zeldovich and A.D. Sakharov, Yad. Fiz {\bf 4}(1966)395; 
Sov. J. Nucl. Phys. {\bf 4}(1967)283.}

\bibitem{Postcard}{
A. D. Sakharov, private communication; H.J.~Lipkin, Annals 
NY Academy of Sci. {\bf 452}(1985)79,
and London Times Higher Education
Supplement, Jan. 20,1984, p.~17.}

\bibitem{DGG}{A. De Rujula, H. Georgi and S.L. Glashow, Phys. Rev. D12
(1975) 147}

\bibitem{clodudek}{J. J. Dudek and F. E. Close, 
arXiv:hep-ph/0311258}

\bibitem{WA89}
J. Pochodzalla, talk at the 2nd Panda Workshop,
Frascati, 3/18-19,2004, "Pentaquarks - facts and mysteries",
\hfill\break
{\tt
www.lnf.infn.it/conference/2004/Panda/Frascati2004\_final\_pochodzalla.pdf}

\bibitem{CDF}
I. Gorelov, talk at DIS-2004, \v Strbsk\'e Pleso, Slovakia, 14-18 April
2004,
{\tt www.saske.sk/dis04/talks/C/gorelov.pdf}\ .

\end{thebibliography}
\end{document}